# EVALUATION OF HIGH SCHOOL PROGRAMME FOR GIFTED PUPILS IN PHYSICS AND SCIENCES IN SERBIA
## - EXPERIENCE IN REGIONAL COOPERATION - SEENET-MTP NETWORK -

G.S. DJORDJEVIC [1], D. PAVLOVIC-BABIC [2], J. STANKOVIC[1]

[1]*Department of Physics, University of Nis, P.O.Box 224, Nis, Serbia*

[2]*Institute of Psychology, Faculty of Philosophy, Cika Ljubina 18-20, Belgrade, Serbia*

Abstract. The "High school class for students with special abilities in physics" was founded in Nis, Serbia (http://tesla.pmf.ni.ac.rs/f_odeljenje/) in 2003. The basic aim of this project has been introducing a broadened curriculum of physics, mathematics, computer science, as well as chemistry and biology. Now, eight years after establishing of this specialized class, we present analyses of the pupils` skills in solving rather problem oriented test, as PISA test, and compare their results with the results of pupils who study under standard curricula. Also, an external evaluation conducted more recently, shows that Special physics class students performed higher on science knowledge test in comparison with students from control groups (grammar school and special math class students). Establishing of the Special physics class as an interesting educational experiment and its development has been connected, in a sense, with activities of the Southeastern European Network in Mathematical and Theoretical Physics. We present the main achievements of the Network and their possible impact to the students. We make conclusions and remarks that may be useful for the future work that aims to increase pupils` intrinsic and instrumental motivation for physics and sciences, as well as to increase the efficacy of teaching physics and science.

Key words: Physics Education, Pupils with Special Abilities, PISA Testing, External evaluation

## 1. INTRODUCTION

In the countries of former Yugoslavia, classes for students with special abilities have a long and successful tradition. Despite some improvement that has been made through last few years, today three fundamental problems still characterize teaching in schools in Serbia: obsolete equipment, obsolete education concepts and insufficient motivation of teachers (for instance - small payroll). One of the consequences of this situation is a very small number of students in natural sciences and engineering sciences at many universities. Moreover skills of pupils in using methods and tools developed in physics and other sciences seem to decrease at the same time with a new revolution in science and technology in developed countries.

The main goals of the project "GRAMMAR SCHOOL CLASS FOR STUDENTS WITH SPECIAL ABILITIES IN PHYSICS", started in Nis, Serbia, in 2003, are to offer a high-quality education, to give gifted pupils a perspective for continuing with high-quality education and to convey initiative and enthusiasm [1, 2]. These goals are to be reached by the following measures: (i) focus on the natural science, on physics in particular, (ii) provision of basic laboratory equipment and PCs (virtual experiments and Internet access), (iii) close collaboration with the University (Host teaching by docents, assistants and project guests, mentors from the university), (iv) close collaboration with similar projects in EU and Eastern Europe, (v) more intensive foreign languages teaching (especially English), for details see [1, 2].

The authors of the curricula and project have faced a lot of problems in implementation of the project in its basic form. However, one of the most important aims has been permanent evaluation of the pupils in the "new class" and comparison of their results with pupils educated in the standard and "mathematical" classes in Serbia. In this paper we present the main results of about 20 successive testing in the period 2003-2010, with more than 4000 individual tests processed, in a brief form.

## 2. EVALUATION IN PHYSICS (2003-2006)

Let us remind, briefly, on the main results in the previous phase of the continuous evaluation of the "Special Class", i.e. Curricula.

The very first testing was done at the very beginning of this ``educational experiment``, early autumn 2003 [1-3]. We will start our presentation with results of two tests in physics made in October 2005 and May 2006, i.e. at the beginning and at the end of the School year 2005/2006, given in the Table 1. There have been five groups of pupils (1.Special class for "physicists" in Nis (9 pupils), 2.Special class for "mathematicians" in Nis (7), 3.Standard grammar class in Nis (20), 4.Special class for "mathematicians" in Belgrade (17) and 5.Special class for "mathematicians" in Novi Sad (7)). All pupils worked out the same test with 20 questions (in total 100 points) and 2 problems (in total 50 points). At this stage we measured the abilities of pupils only in physics and mainly in the first grade of high school. The differences in syllabus in physics are so big in the second and third year that comparison of results is sensible just after the end of the grammar school, i.e. after 4$^{th}$ year.

Let us focus on results of the third generation - pupils born in 1990. The other results were presented in the two reports of the Special team established by the Serbian Ministry of education in 2005 and 2007. In the first column we denote the corresponding class and the numbers of pupils that took part in both testing. In the following columns one can see their records in test the questions and solving problems in percents.

Table 1
Results of the third generation – born in 1990

| Level of complexity | Quest1 (%) | Prob.1 (%) | Tot. 1 (%) | Quest2 (%) | Prob.2 (%) | Tot. 2 (%) |
|---|---|---|---|---|---|---|
| "Physicists"-Nis (9) | 52,67 | 22,4 | 42,58 | 71,11 | 27,78 | 56,67 |
| "Mathematicians"-Nis (7) | 35,00 | 0,00 | 23,33 | 64,86 | 4,29 | 44,67 |
| Standard class-Nis (20) | 42,60 | 0,00 | 28,40 | 61,40 | 0,00 | 40,93 |
| "Mathematicians"-BG (17) | 67,06 | 0,71 | 44,94 | 68,29 | 29,18 | 55,25 |
| "Mathematicians"-NS (7) | 69,14 | 8,29 | 48,86 | 81,86 | 31,43 | 65,05 |

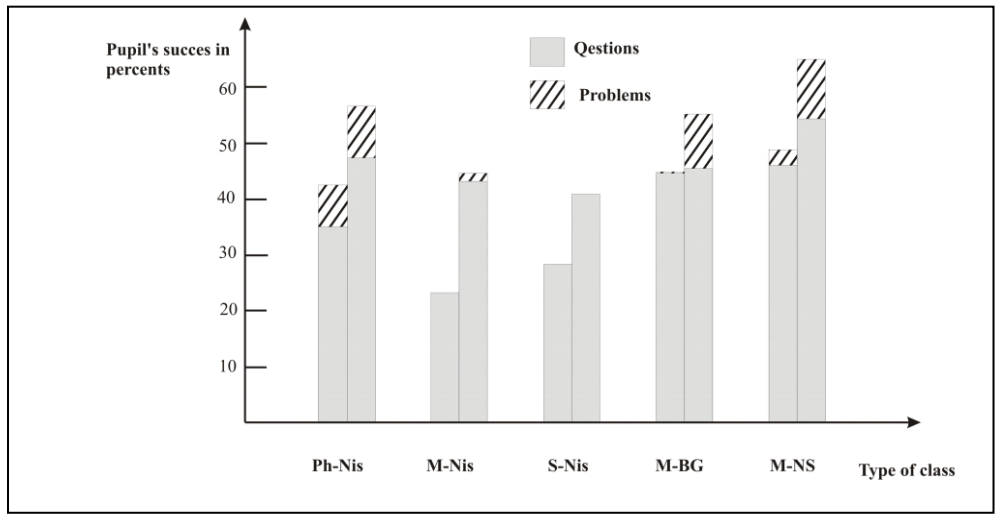

**Figure 1 - Pupil's success in solving test questions and problems, pupils born in 1900. Ph-Nis, M-Nis, S-Nis, M-BG, M-NS.**

From results presented in Table 1, i.e. Figure 1, as the most instructive, when a system of selection of pupils was established and properly implemented, we can conclude several things: (i) ``physicists`` showed significantly better abilities in physics, in particular in solving problems than two classes of high school students from Nis and the same or slightly better abilities than high school students from the elite Mathematical Gymnasium in Belgrade. Results and statistics from Novi Sad were based on a small group and some later testing indicated that sample of (7) pupils were not representative of the whole class, but anyway it does not change the main conclusions, (ii) results of ``physicists`` at the second test shows a stabile and good improvement in their abilities in both – questionnaires and solving problems aspects.

## 3. EVALUATION IN PHYSICS (2006-2008)

There were several tests in physics during that period (2006-2008) as a continuation of testing in the previous period [4]. It should be noted that every of the first 4 generations of "physicists" has been followed and evaluated at least twice per year, as well as the corresponding group of "mathematicians" and at least one "control" group of pupils in a Grammar school - "standard Curricula", mostly in Nis. Presentation of the contemporary results and analysis would need much more "room" than we have at our disposal in this paper. However, in this section we will present results concerning the first generation of "physicists" and groups of pupils we evaluated permanently through their 4 years education it the Grammar schools.

4Let us consider results of two independent batteries of tests created by different groups. An independent group of evaluators nominated by the Serbian Ministry of Education both made in spring 2007. The first test was prepared by the external team of the evaluators. There were four groups of questions and problems in physics, with increasing complexity, from 1 to 4. The results are presented in the Table 2. All pupils were in the 4$^{th}$ grade and they took part in the first testing, as well as in all other evaluations started in September 2003.

Table 2
Results of the first generation (born in 1988), Test No1 spring 2007 – 4$^{th}$ grade

| Level of complexity | 1 (%) | 2 (%) | 3 (%) | 4 (%) | Total (%) |
|---|---|---|---|---|---|
| "Physicists"-Nis (9) | 93,3 | 95,6 | 69,4 | 69,66 | 85,68 |
| "Mathematicians"-Nis (8) | 94,6 | 89,7 | 71,1 | 37,5 | 74,26 |
| Standard class-Nis (25) | 82,4 | 23,3 | 4,7 | 0,0 | 34,47 |

The second test was prepared by the permanent group of evaluators form Nis. This test was similar to the previous ones, slightly improved and also with classified problems in 5 categories. The results were summarized in Table 3.

Table 3
Results of the first generation (born in 1988), Test No2 spring 2007 – 4$^{th}$ grade

| Level of complexity | 1 (%) | 2 (%) | 3 (%) | 4 (%) | 5 (%) | Total (%) |
|---|---|---|---|---|---|---|
| "Physicists"-Nis (7) | 100,0 | 87,2 | 58,33 | 100,00 | 71,43 | 37,93 |
| "Mathematicians"-Nis (13) | 84,62 | 47,92 | 32,05 | 43,59 | 25,64 | 19,81 |
| Standard class-Nis (11) | 72,73 | 46,78 | 20,45 | 3,03 | 0 | 14,32 |

Numbers in the small brackets denotes numbers of the pupils form the corresponding group who took part in the particular evaluation. Despite a relatively small number of pupils and fluctuation of some pupils, when the absence of a top or bottom achievers from the group can change the total score, it is obvious that "physicists" after four years spend in the special class with the new curricula have excellent results, and that their abilities and skills in physics are significantly better than "mathematicians'", who were selected at the beginning of high school study. ("Physicists" in this generation were not because they could enroll the class only on the basis of their wishes). Let us remind ourselves of the results of the same classes at the very first testing.

Table 4
Results of the first generation – 1$^{st}$ grade

| Level of complexity | 1 (%) | 2 (%) | Total (%) |
|---|---|---|---|
| "Physicists"-Nis (17) | 28,6, | 4,47 | 21,69 |
| „Mathematicians"-Nis (16) | 34,69 | 7,06 | 27,83 |
| Standard class-Nis (27) | 32,01 | 0,00 | 21,34 |

Even on the first glance it is visible that ``physicists`` has doubled their score after 4 years in comparison to ``mathematicians`` and that results in respect to the ``standard class`` are even better. These results, beside excellent results of pupils form this ``special`` class in (inter)national contests in physics, mathematics etc, were of the crucial importance for the Special team of evaluators to propose to the Ministry of education transformation of the Class` status from an ``experiment`` to a standard educational profile in Serbia.

### 4. PISA TEST

Trying to collect more valuable and comparative data about achievement of pupils and achievement correlates, we included OECD/PISA (Programme for International Student Assessment) science competencies tests in the evaluation schema. The main intent was to establish a model of monitoring and external evaluation of achievement focused on functional knowledge, which enable us a) to compare the quality of knowledge in domain of physics and, even more, of science, to internationally recognized criteria (key competencies); b) to document the well-being of schooling in special teaching program, e.g. comparing their achievement with classic course, and c) to understand in what extent higher order thinking is supported by different types of schooling.

PISA 2006 defines scientific literacy in terms of an individual's scientific knowledge to identify questions, to acquire new knowledge, to explain scientific phenomena, and to draw evidence-based conclusions about science-related issues. In addition, pupil shows the understanding of the characteristic features of science as a form of human knowledge and enquiring. For detailed description see [5]. In other words, PISA is not limited to measure pupil's acquisition of curricula or specific science content, but his/her capacity to identify and understand science related issues as well as the capacity to interpret and apply science evidence in order to solve problems and make decisions in real-life situations [6].

### 4.1. Sample

All pupils in special classes for physics in the $4^{th}$ and the $1^{st}$ grade (11, 10 pupils, respectively) were included in testing. In addition, in order to compare data, special math class grade $4^{th}$, Grammar school the $4^{th}$ and the $1^{st}$ grade, were included as well (14, 11, 32 pupils respectively). Also, we use PISA 2006 national database to compare achievement data.

We are aware that all conclusions we drawn, based on comparison between this and national sample, are uncertain due to small size of the selected sample, as well as to the age difference (PISA pupils are mostly $1^{st}$ grade pupils). But, this investigation is still valuable in our case because: we have a tool to measure pupil`s skills not just in physics than more generally in science and solving problems; PISA is a widely accepted and procedure with clear standards and reliability and validity.

Testing took part at the end of the school year (May 2007), i.e. at the beginning and at the end of the secondary schooling. Every student was given a 2-hours paper-pencil science tests.

### 4.2. Results

*Science Competencies*

The science scale was constructed to have a mean score among OECD participating countries of 500 points, and standard deviation of 100 points, which means that it is expected to have about two thirds of pupils scoring between 400 and 600 points. Table 4 shows the average achievement of OECD countries pupils and Serbian pupils in PISA 2006.

Table 5
Average performance – PISA 2006

| Referent sample | Mean |
|---|---|
| OECD countries | 500 |
| Serbia, the whole sample | 436 |
| Serbia, Grammar school pupils | 501 |

In short, results show us that, as a whole, Serbian education system performs under the expected average. Knowing that one year of schooling adds about 40 points on this scale, we can say that Serbian education has lost (or has spent on nothing) one and a half school year out of 9 years (which is the duration of formal schooling of PISA pupils in the moment of testing). Grammar school pupils are at the level of their OECD counterparts, but this is the whole OECD sample, which includes all groups of pupils, while in Serbia, in Grammar schools we have pupils positively selected by their school competencies.

Achievement data shows that "physicists" grade $4^{th}$ is slightly above the average achievement, but here we have 3 school years older pupils. On the other hand, the achievement of two remaining special classes is far above the average. Those results are respectful even knowing that we are talking about two relatively small groups of pupils. Additional analysis helped us to understand better the nature of these outstanding results.

*Science Competencies: Levels of Achievement*

Pupil performances on science are grouped into six proficiency levels. This grouping was taken on the basis of substantive considerations relating to the nature of the underlying competencies. Each pupil is positioned in the highest level in which she or he can answer the majority of tasks. So, levels 1 and 2 show what is the critical baseline for science competencies. On the top of the scale (proficiency levels 5 and 6) we have pupils enabled to solve complex problems, related to scientific knowledge and understanding scientific data, which require several interrelated steps, as well as the use of critical thinking and abstract reasoning [6].

Following graphs show some comparative data about the distribution of pupils in these two top levels of performance.

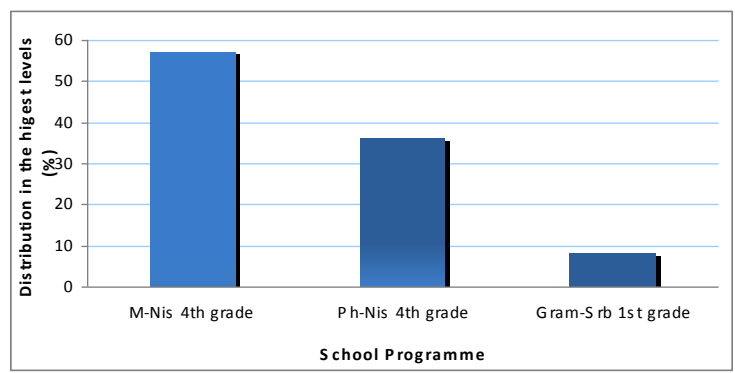

**Figure 2 - The percentage of pupils in the highest levels of performance (4$^{th}$ grade – the first generation – born 1988 - gifted pupils and PISA Serbia Grammar school sample)**

Figure 2 shows us that high performance is very rare among 1$^{st}$ grade pupils of Grammar schools in Serbia, less than 10% of them can successfully solve the most complex PISA tasks. At the same time, more than half of gifted mathematicians are able to perform on these levels. But, the most interesting data is referring to the distribution of performance in the group of "physicists". While this group of pupils performs, in average, at the same level as the Grammar school students, here we can see higher concentration of top performances in this group than in Grammar school group (almost three times more). Still, we should keep in mind the fact that we speak about pupils of different ages and, consequently, of different length of schooling, as well as the very small groups of pupils in special classes. It should be also noted that this first generation of ``physicists`` was very specific one, with about one half (8 or 9) of pupils enrolled in the first round and with a quite good record from their elementary schools, and rest of the pupils, up to final 15 who finished the high-school was enrolled in the second round with a record much lower of the first part. Because of that, results of the next, in particular the third, fourth etc, generations are much more illustrative and important.

So, let us see what the result of comparison between the 1$^{st}$ grade gifted pupils and their counterparts in Grammar schools are (Figure 3).

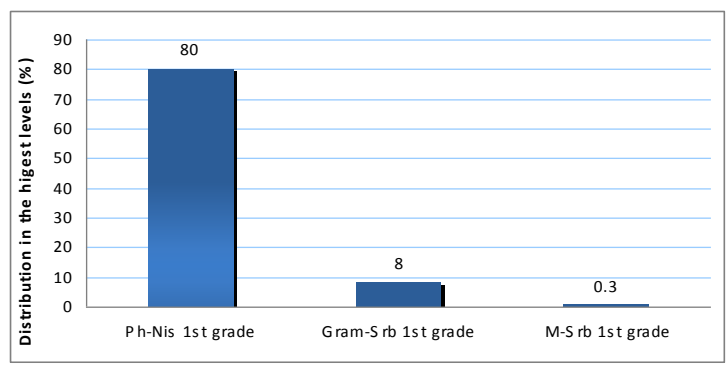

**Figure 3 - The percentage of pupils in the highest levels of performance (the 1st grade – gifted pupils – the fourth generation – born 1991, PISA Serbia Grammar school and PISA Serbia whole sample).**

As we can see, almost all pupils in this class have the highest ranking scores. Talking in language of competencies, these pupils demonstrate advanced scientific thinking and reasoning, and they are able to solve complex and unfamiliar scientific or technological situations.

We can draw some policy-oriented implications on these results. First of all, it is obvious that groups of gifted pupils, especially those in 1st grade, deserve and/or need very carefully designed curricula which will engage their already built competencies, as well as enable them to develop their capacities even more. Also, it is very important for them to have well-trained teachers sensitive enough to follow their educational needs. Then, we can say that demonstrated thinking skills have transfer effects on other (related) domains, so we can expect them to be high achievers in other subjects, as well [7].

## 5. EXTERNAL EVALUATION

Lastly, in 2009 the Ministry of Education of Republic of Serbia, as a founder of the Programme, initiated a process of Programme evaluation. The aim was "to test the level of compliance of predefined goals and outcomes of the Programme". The evaluation was conducted by an independent government agency, the Institute for Education Quality in Belgrade. The evaluation report was finsidhed and published in March 2010 [8].

### 5.1 Metodhology of Evaluation

It is possible to distinguish three main components of evaluation, based on different data types:
1. Student achievement on a science knowledge test (taken by 3 groups of students/high-school pupils: basic group – students of all grades involved in the Programme, and two control groups – students of Grammar school and students involved in special math classes);

2. Perceptions of students, teachers and school management about the planned solutions, applied practices and the quality of teaching, as well as the possibilities and limitations of the Programme in the implementation of the curricula; and
3. Content analysis of existing documentation.

In this paper, we will present only the findings based on quantitative methodology (the achievement test and a questionnaires examining perceptions and attitudes of students and teachers), because these findings are most relevant to understanding the quality of the achievements of students involved in the Programme.

### 5.2. Sample

Student sample: All pupils in special classes for physics (Programme) from the 1$^{st}$ to the 4$^{th}$ grade were included in testing (in sum, 36 students). In addition, in order to compare data, special math class grade 1$^{st}$ to 4$^{th}$, and Grammar school students grade 1$^{st}$ to 4$^{th}$ were included as well (38, 47 students respectively).

Teacher sample: in this subsample were included 12 teachers and 5 representatives of school management.

We are aware that all conclusions, based on comparison between different groups, we drawn are very sensitive and, possibly, are uncertain in a part due to small size of the selected subsamples.

### 5.3. Main findings

*Achievement on Knowledge Test*

A science knowledge test were given to all sampled students. In construction of the test, publicly available tasks taken over from two international assessment projects, TIMSS and TIMSS Advanced, were used. In terms of cognitive complexity, the test is composed mainly of tasks which perform on higher levels of achievement (in other words, more difficult tasks were predominant). The following cognitive domains were represented in the test: knowledge, application and reasoning. In terms of content, selected tasks were in mathematics and physics. The pilot testing has shown that the internal consistency of the test is .741, which makes it acceptable [8].

Testing the differences in means shows that Special physics class students have achieved significantly higher scores on knowledge test then Grammar school students and equally good achievement as Special math class students (Figure 4). When looking only tasks in physics, it was found that the **Special physics class students achieved statistically higher results in comparison to both control groups of at least 22%.**

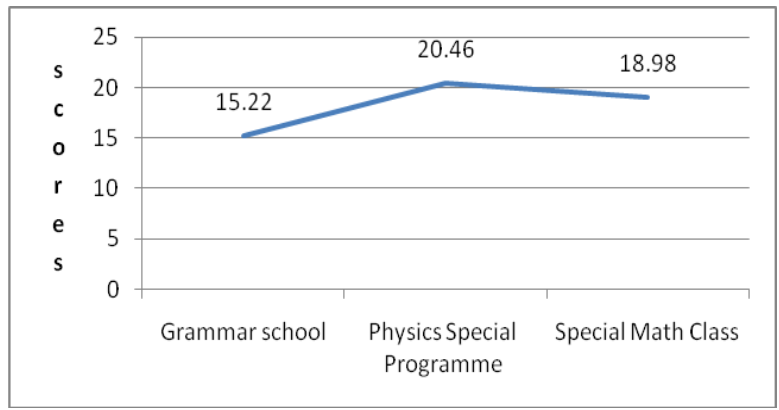

**Figure 4 - Mean scores on science knowledge test: all groups of students**

Analysis of knowledge test means by grades shows that there is a continuous improvement in achievement for Special physics class students, while this does not apply to students from control groups (Figure 5).

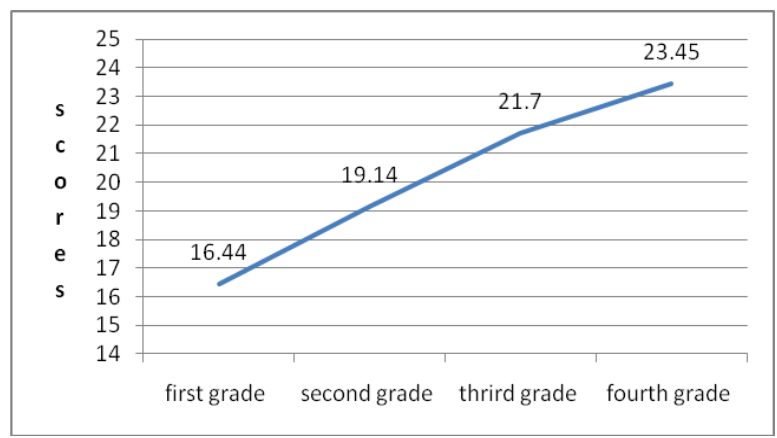

**Figure 5 - Mean scores on science knowledge test by grades: Special physics class students.**

*Perception of Students and Teachers*

Students of all programmes, as well as Special physics class' teachers filled out questionnaires with parallel thematic structure. The main thematic areas that appear in the questionnaire were: curriculum, assessment, teaching and extracurricular activities, teacher-student relationship and competition.

Special physics class students give more positive marks for the quality of the programme they are involved in, in comparison with both control groups of students. They claim to have significantly more opportunities for research, project designing and conducting, as well as for the experimental work. They estimate that they are better prepared for further education then students from control groups. Also, there is agreement between Special physics class students and teachers that the curriculum is based on relevant content, but it is too large.

A significant finding relates to the perception of students about the quality of the relationship that exists between teachers and students. Special physics class students have given statistically significantly higher grades for all investigated dimensions of this relationship (between 4.7 and 4.8 of 5.0(!) in comparison with 3.0 to 3.8 of 5.0 in other classes in this testing). Specifically, they estimate that their teachers in a greater extent support the independence of students and encourage their active participation in class, they respect students' creativity and ideas and they treat them with respect.

## 6. THE SEENET-MTP NETWORK, A "BRIDGE" BETWEEN HIGH-SCHOOLS, HIGHER EDUCATION AND RESEARCH

Recognizing the importance, as well as the necessity, of bridging the gap between Southeastern and Western European scientific community, the participants of the UNESCO-ROSTE - sponsored BALKAN WORKSHOP BW2003 "Mathematical, Theoretical and Phenomenological Challenges Beyond the Standard Model: Perspectives of Balkans Collaboration" (Vrnjacka Banja, Serbia, August 29 - September 3, 2003, http://www.pmf.ni.ac.rs/pmf/bw2003/index.html) came to a common agreement on the Initiative for the SEENET-MTP NETWORK. Within the years 2004-2009, 17 institutions from 8 countries (Bosnia and Herzegovina, Bulgaria, Croatia, Greece, Macedonia (FYROM), Romania, Serbia, Turkey) in the region joined the Network (http://www.seenet-mtp.info/network-nodes.html), 11 partner institutions all over the world, as well as about 250 individual members with an increasing trend. This Network has been a natural continuation of Prof. Julius Wess` initiative: Scientists in Global Responsibility, started in 1999. In accordance with the decisions made at the last meeting of the Representative Committee of the Network (April 2009), president of the Network is Prof. Dr. Radu Constaninescu (University of Craiova, Romania), Executive director of the Network and its permanent office in Nis is Prof. Dr. Goran Djordjevic (University of Nis, Serbia), Coordinator of the Scientific-Advisory Committee of the Network is Prof. Dr. Goran Senjanovic (ICTP, Trieste, Italy). The activities of the Network are closely connected to and indirectly support other efforts to promote physics and the study of physics in the region, such as the Special physics class in Nis.

One of the most complex project implemented by the Network is preparation of "The Map of the top research groups and organizations in the fields of Physics and Mathematics", following their publications records from 2000-2009, with a reach collection of data from 42 Institutions from 7 countries in the region. Beside research and training program for the PhD students, as one of the most important reasons for establishing of the Network and its main activity [9], what is not a topic we will consider in this paper, a permanent activity has been toward: "Promotion of the excellence and growing of the youth's interest for education in Physics (and Mathematics),", evaluation of the ``Special Class in Physics``, joint regional contests in Physics and popular lectures mainly organized in Serbia and Romania.

Let us mention some numeric facts, even we did not testing and could not present some evaluation of this part of the program now, as we did for the high-school class for ``physics and science``, it could be helpful in understanding level of activities done in the last 8 years.

The Network, through its nodes, organized 11 scientific meetings, mainly in Serbia, Romania and Bulgaria, with about 750 participants. There were about 150 short-term scientific exchanges and visits, and 5 mid-term fellowships were provided for undergraduate and graduate students. Of a special importance were **Three International Meetings and Contests "Science and Society"** organized in Romania (in total with more than 120 competitors). The last one took place form 15 to 17 April 2011, in Turnu Severin, Romania. More than 40 pupils from Romania and Serbia took part as well 30 teachers from the same countries. Pupils-competitors have shown excellent abilities in solving problems, and comparison of results of pupils from the ``special class`` from Nis from 2008 to 2011, shows a continuous increasing trends in their relative and absolute results (http://www.seenet-mtp.info/news/the-third-edition-of-the-international-meeting-science-and-society-results/) . Beside a honorably mention of Vladan Pavlovic at the Physics Olympiad in Iran, this conclusion could be also accepted as confirmed by the bronze medal of Tamara Djordjevic at Olympiad in Thailand, most recently.

More than 150 researchers, students and teachers in the four successive meetings in follow, as a integral part of the **BALKAN SUMMER INSTITUTE - BSI2011** from August 19 – September 1, 2011, Nis and Donji Milanovac, Serbia will consider possibility to continue this program, those results presented in this paper and new modalities and forms of action in the next 2-4 years

## CONCLUSION

Program of evaluation of the curricula and its implementation in this educational experiment in Serbia, even though it was focused most on physics, at this stage is quite nontrivial. Physics syllabi differ across schools and classes, the similarities are significant only in the first year and after that rather different until the end of the grammar school when they have the same "core" in physics. Together with a new approach in teaching resulted in a better score of pupils of at least 22%.

It was found that pupils from the standard class are not able to solve problems (their records in solving problems tend to zero in almost all generations). The "physicists" show slightly better improvement in physics, and continually good records in solving problems. It is worth to note that the new class and program "for physicists" has attracted better pupils and that number of pupils is increasing 7, 11 and 15 and nowadays is stabilized around 16 per generation. It can be explained by attractive curricula, a lot of guest lecturers, additional laboratory work, excursions, some support in books and awards. The results of former pupils of this class, current students at universities all over Serbia but also at MIT, LMU

Munich etc. are excellent as well as their perception of quality of curricula, teaching in the ``Special`` class and benefits for pupils in their future education [8].

Finally, the students and teachers perceptions strongly recommend the revision of curricula in order to reduce the range of selected contents, but keep the achieved level of quality, as well as the performance level.

## ACKNOWLEDGMENTS

This work and evaluations has been partially supported by UNESCO-BRESCE grants No. 8758346, 8759228, UNESCO-Office in Venice No. 4500143843, ICTP grant PRJ-09 as a part of the support to the SEENET-MTP program, and by the Projects of the Serbian Ministry for Education and Science No. 179018 and No. 176021. We would like to thank all the colleagues from Nis, Belgrade and Novi Sad who take part in this seven years long period of permanent evaluation. We are exceptionally thankful to Lj. Nesic, T. Misic and Lj. Kostic-Stajkovic, for preparation of many tests and their evaluation. G. Dj would like to thank to R. Constantinescu for a great effort and excellent organization of three bilateral Romanian-Serbian contests and warm hospitality. G.Dj would also like to thank M. Visinescu, R. Constantinescu, D. Vulcanov and I. Cotaescu for the crucial support in establishing the Network and many excellent lectures given both for the high-school pupils, undergraduate and graduate students, first of all in Nis. It is our pleasure to thank D. Dimitrijevic for their continual support in preparing test materials and their realization.